\documentclass{biophys_letter}
\usepackage{dcolumn,color,helvet,times}
\usepackage{amsmath,lastpage,bm,textcomp}%mathtime
\usepackage{multicol}
\usepackage{graphicx}
\jno{kxl014} %journal number
\gridframe{N}%option for grid around the text "Y" or "N"
\cropmark{N}%option for cropmark around the text "Y" or "N"

\doi{doi:}% DOI number in the copyright line

\begin{document}

%\onecolumn

\setcounter{page}{1} %first page number
\title{Size-dependent rheology of type-I collagen networks}

\author{Richard C. Arevalo, Jeffrey S. Urbach and Daniel L. Blair}

\address{Department of Physics, Georgetown University, Washington, DC,
  20057, USA}

\maketitle

\pagestyle{headings}

\markboth{Biophysical Journal: Size Dependent Rheology}{Biophysical Journal: Size Dependent Rheology} %for running head

\begin{abstract}
{We investigate the system size dependent rheological response of
  branched type I collagen gels. When subjected to a shear strain, the
  highly interconnected mesh dynamically reorients, resulting in
  overall stiffening of the network. When a continuous shear strain is
  applied to a collagen network, we observe that the local apparent
  modulus, in the strain-stiffening regime, is strongly dependent on
  the gel thickness.  In addition, we demonstrate that the overall
  network failure is determined by the ratio of the gel thickness
  to the mesh size.  These findings have broad implications for
  cell-matrix interactions, the interpretation of rheological tissue
  data, and the engineering of biomimetic scaffolds.}{Received for
  publication XX and in final form XX.}  {Address reprint requests and
  inquiries to D.~L.~Blair, E-mail: blair@physics.georgetown.edu.}
\end{abstract}

\vspace*{2.7pt}
\begin{multicols}{2}
The increasing variety and availability of purified extra- and
intra-cellular matrix proteins provides an unprecedented opportunity
to quantify the mechanical properties of the cellular
environment. Reconstituted biopolymer fiber networks can exhibit
nonlinear rheology that is dramatically different than synthetic
polymer gels
~\cite{gaill_biorheo_93,weitz_04,janmey_nature_05,liu_prl_07,fletcher_nat_07}.
Tying the microscopic physical behavior of the network constituents to
the meso- and macro-scale rheology is critical for determining the
mechanosensory mechanisms that underlie such diverse processes as cell
motility, cancer metastasis and tumor proliferation. {\em In vivo},
cells sense, and respond to forces generated within or transmitted
through the extracellular matrix (ECM). In turn, the regulation of
sensory cues can be highly susceptible to changes in the ECM stiffness
on length scales comparable to the cell size. However, in spite of the
increasing number of experimental and theoretical investigations of
the nonlinear rheological behavior of semi-flexible and stiff polymer
networks ~\cite{janmey_prl_95,frey_prl_03,head_pre_03,frey_pre_07},
the effects of system size on strain-stiffening remains essentially
unexplored. Most theoretical models focus on the global response of
biopolymer networks using three distinct microscopic mechanisms;
non-affine deformations of crosslinked rigid rods~\cite{frey_pre_07},
entropic penalties of fiber stretching in entangled
networks~\cite{janmey_nature_05}, or changes to the network
geometry~\cite{onck_prl_05}.  There is little experimental data to
directly verify these microscopic mechanisms, in part because of the
difficulty in extracting the relevant quantities
experimentally~\cite{stein_micro_08,kaufman_biophys_09}.  In addition,
the nonlinear rheology of stiff networks can only be investigated with
bulk techniques, so the behavior at biologically relevant microscopic
and mesoscopic scales is largely unknown.

\begin{figure}[tb]\vspace*{-5pt} %% Figure 1
\centering{\includegraphics[width=0.5\textwidth]{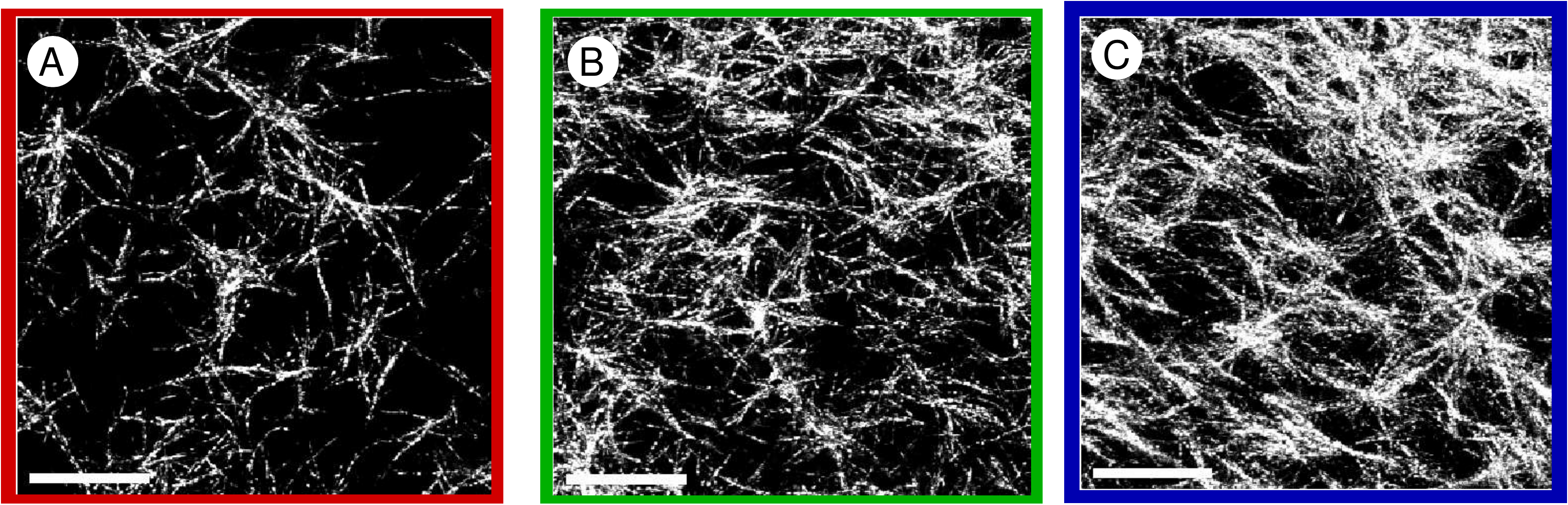}}\vspace*{-12pt}
\caption{\label{images}  Confocal reflectance images of branched type I rat tail
  collagen fiber networks corresponding to the concentrations: ({\bf
    A}) 1 mg/ml, ({\bf B}) 2 mg/ml, ({\bf C}) 3 mg/ml. Each image is a
  10 $\mu$m thick maximum-projection; scale bar = 25 $\mu$m.}\vspace*{-18pt}
\end{figure}

Collagen is the most abundant ECM protein, and a common component of
{\em in vitro} cell cultures and bioengineered scaffolds.  Under
appropriate conditions, {\em type I} collagen self-assembles {\em in
  vitro} to form percolated networks (gels)~\cite{sackmann_biophys_03}
with mesh sizes $\xi$ that depend on the concentration and
polymerization conditions (Fig.~\ref{images}A-C)
~\cite{kaufman_05,kaufman_biophys_09}. These branched networks
substantially strain-stiffen over a broad range of
concentrations~\cite{janmey_natmat07,mackintosh_09, vader_plosone_09};
one interpretation of the biomechanical function associated with
nonlinear stiffening is a passive protection for soft tissues when
shear deformations become large. Collagen networks are not only
excellent ECM model systems, but provide a platform for investigating
cellular motility within three-dimensional microenvironments
\cite{fabry_08}, and provide bio-compatible scaffolds for tissue
growth and organ regeneration
\cite{taylor_08,margaret_08,wang_02,wong_03}.

In this work, we investigate strain-stiffening of collagen networks
under steady shear and observe that the nonlinear rheological response
is strongly dependent on the material thickness. Moreover, the
apparent moduli near yield decreases dramatically in thin samples in a
process that is controlled by the ratio of the sample size to the
network mesh size. The system size dependence is not accounted for in
current models of nonlinear strain-stiffening in biopolymer networks.

%Unleashing the full potential of these applications requires an
%understanding of the connection between the nonlinear macroscopic
%rheology and the network microstructure. .

\begin{figure}[b!]\vspace*{-5pt} %% Figure 2
\centering{\includegraphics[width=0.33\textwidth]{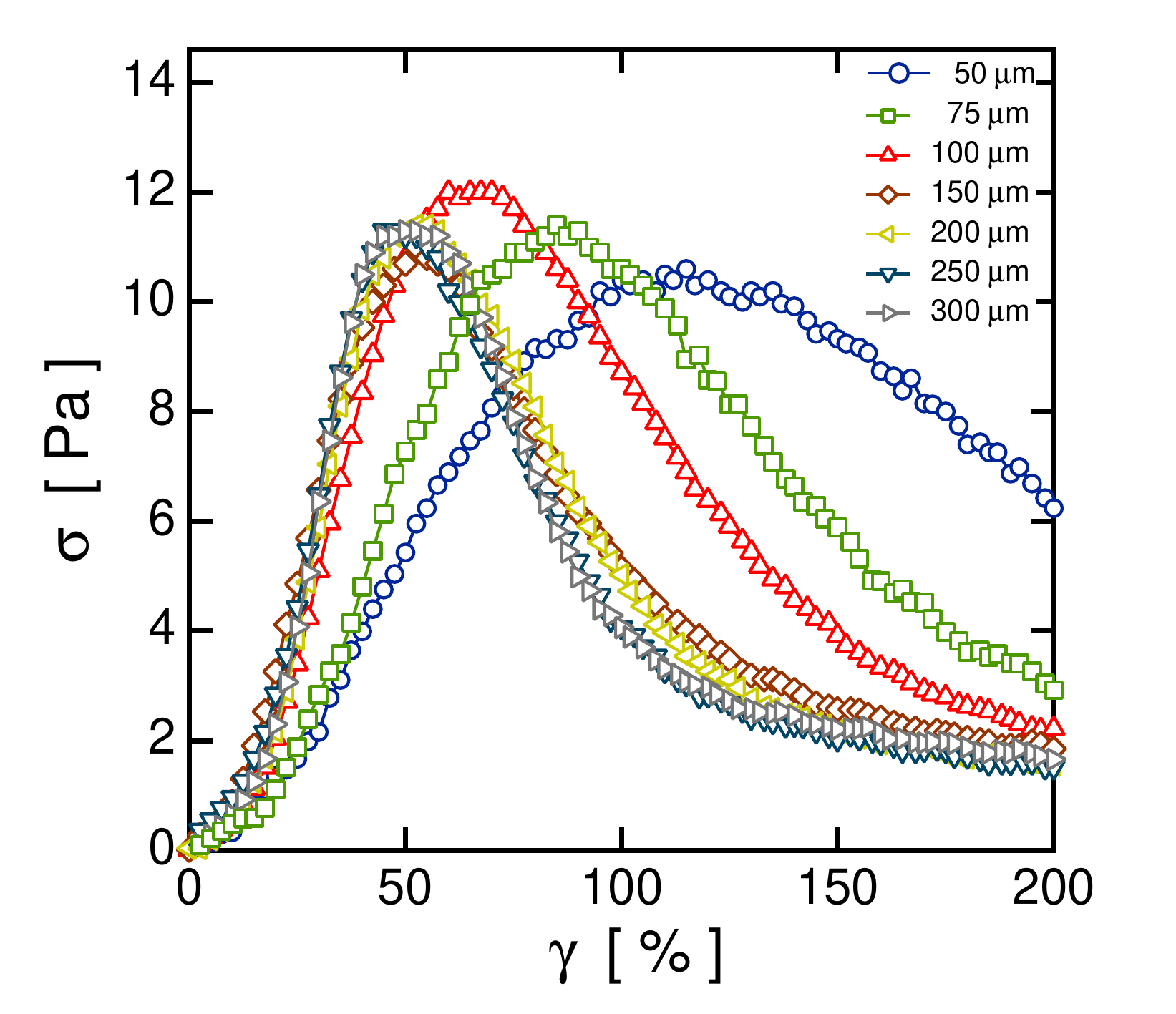}}\vspace*{-16pt}
\caption{\label{strstr} Stress-strain rheology curves for 1 mg/ml
  collagen gels at different gaps $h$ (strain rate = 1\% s$^{-1}$);
  gel networks are polymerized at $23^o $C and pH 7.0. All lines
  connecting points are guides for the eye.}\vspace*{-15pt}
\end{figure}

\begin{figure}[tb]\vspace*{-5pt} %% Figure 3
\centering{\includegraphics[width=0.3\textwidth]{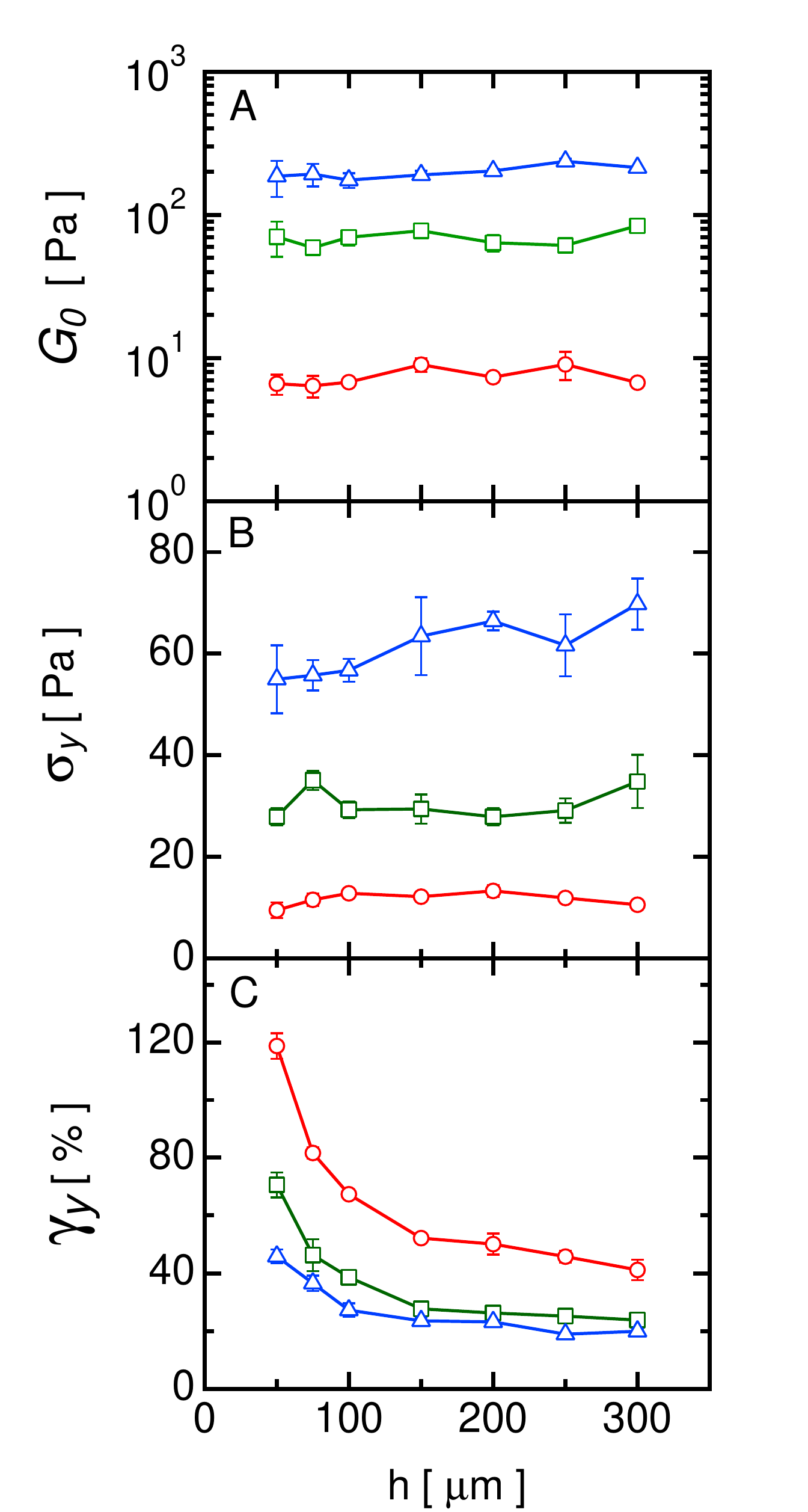}}\vspace*{-13pt}
\caption{\label{g0_yields} For all graphs, ($\circ$) 1 mg/ml, ($\Box$)
  2 mg/ml, ($\triangle$) 3 mg/ml. ({\bf A}) Linear elastic modulus
  $G_0$ from oscillatory rheology versus gap height $h$. ({\bf B})
  Yield stress extracted from the stress-strain curves $\sigma_y$
  versus $h$ shows strong $\xi$-dependence and limited $h$-dependence
  for all concentrations. ({\bf C}) Yield strain $\gamma_y$ versus $h$
  for each collagen concentration extracted from the peaks in the
  stress-strain curves (Fig. \ref{strstr}).  Error bars represent
  standard error from four trials, lines are guides to the eye.}\vspace*{-15pt}
\end{figure}

We apply continuous shear strains to collagen networks using an
Anton-Paar MCR-301 bulk rheometer with a 25 mm diameter parallel-plate
geometry within a temperature and humidity controlled environment. All
continuous shear strain experiments are performed at a strain rate of
$\dot{\gamma}= 1.0\%$ s$^{-1}$. {\em Type I} rat tail collagen (BD
Bioscience, San Mateo, CA, 3.27 mg/mL) is polymerized at 23$^o$C for
45 minutes in 10X PBS at pH 7.0 with ionic strengths
$I=\{0.044,0.087,0.13\}$ for the concentrations $c=\{1.0,2.0,3.0\}$
mg/ml, respectively. 
\end{multicols}
\doiline
\twocolumn
\noindent The plate tool provides precise control of the
sample thickness through a change of the rheometer gap $h$.  At low
strains, the measured stress $\sigma$ is proportional to the applied
strain $\gamma$, for all values of $h$. Intermediate strains reveal
that the network undergoes a substantial nonlinear increase of
$\sigma$, indicative of
strain-stiffening~\cite{janmey_nature_05}. Specifically, for $h
>150\mu$m, $\sigma(\gamma)$ is $h$-independent. However, for smaller
gaps, we observe that the maximum of $\sigma(\gamma)$ moves to larger
values of $\gamma$ (Fig.~\ref{strstr}). The magnitude of $\sigma$ at
the peak remains relatively constant. The yield strain in thin samples
is more than twice that of thick samples, and the apparent modulus of
thin samples can be up to 3 times smaller than the bulk value. To
ensure that the changes in $h$ are not introducing variability in the
rheological measurements, particularly at small $h$, we measure the
linear viscoelastic modulus $G_0$ using oscillatory rheology at a
frequency of $\omega = 2\pi$ rads s$^{-1}$ at a strain of $\gamma =
1.0\%$. For each concentration, we observe that $G_0$ is constant over
the entire range of $h$ (Fig.~\ref{g0_yields}A).
Higher strains lead to irreversible deformations and
yielding, as evidenced by a reduction in the overall modulus. In
addition to this distinctive rheological signature, we observe a
dramatic change in the rheological response by varying the sample
thickness from $h=50-300\mu$m, at a fixed collagen concentration
($c=1.0$ mg/ml).

To quantify the $h$-dependence of the nonlinear rheology, we perform a
least squares polynomial fit to the peaks of the rheology curves in
Figure \ref{strstr}A. We use the peak positions to define the {\em
  yield} values for the stress $\sigma_y$ and strain $\gamma_y$.  We
determine $\gamma_y$ as a function of $h$ for three different collagen
concentrations.  For each concentration, $\gamma_y$ exhibits a clear
increase for small $h$ and approaches a constant value for large $h$
(Fig.~\ref{g0_yields}C). Interestingly, for the entire range of $h$
and concentration the values of $\sigma_y$ show weak or no gap
dependence (Fig.~\ref{g0_yields}B). We speculate that $\sigma_y$ is
determined by the strength of fiber-fiber or fiber-boundary junctions,
and depends only on the total applied stress, independent of $h$.

\begin{figure}[tb] %% Figure 4
\centering{\includegraphics[width=0.5\textwidth]{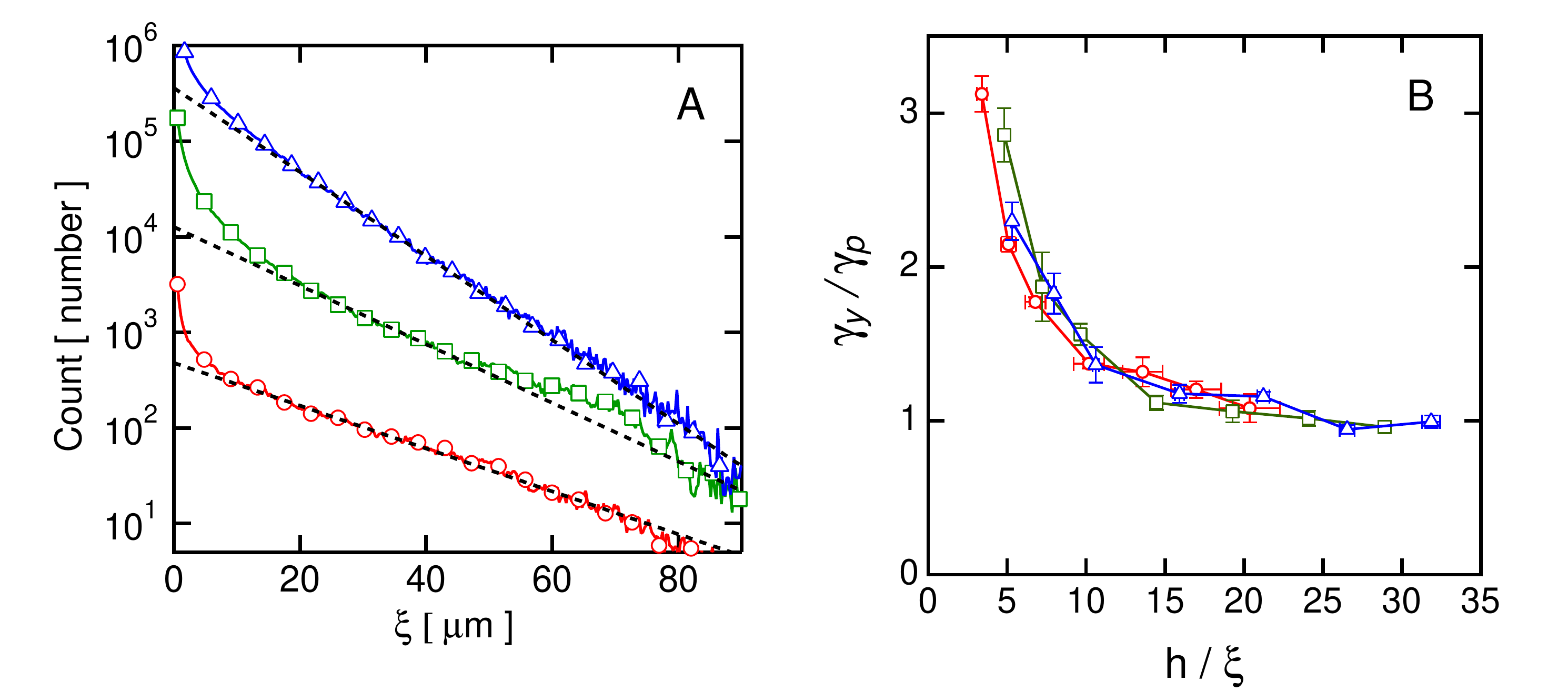}}\vspace{-15pt}
\caption{\label{rescale} ({\bf A}) Mesh size distributions for ($\circ$) 1
  mg/ml, ($\Box$) 2 mg/ml, ($\triangle$) 3 mg/ml. Exponential fits,
  given by the dashed lines, provide characteristic mesh sizes:
  ($\circ$) 14.7 $\mu$m, ($\Box$) 10.4 $\mu$m, ($\triangle$) 9.4
  $\mu$m. These curves have been shifted along the ordinate for
  clarity. ({\bf B}) Rescaling $\gamma_y$ by the extrapolated
  plateau yield strain $\gamma_p$ and $h$ by $\xi$ reduces Figure
  \ref{g0_yields}{\bf C} to a universal yielding curve for all
  concentrations; ($\circ$) 1 mg/ml, ($\Box$) 2 mg/ml, ($\triangle$) 3
  mg/ml.}\vspace*{-22pt}
\end{figure}

To elucidate the role of the microscopic length scale $\xi$ in the
observed size-dependent rheology, we replot the data in Figure
\ref{g0_yields}C with $h$ rescaled by $\xi$.  To quantify the mesh
size $\xi$, we analyze the spacing between fibers extracted along
horizontal and vertical lines within $z$-resolved planar confocal
images \cite{kaufman_05}. We fit the distribution of the fiber spacing
to an exponential decay for each concentration, $c=\{1.0,2.0,3.0\}$
mg/ml, to obtain the characteristic mesh sizes $\xi=\{$14.7, 10.4,
9.4$\}$ $\mu$m (Fig.~\ref{rescale}A). To account for the concentration
dependent modulus, we also rescale $\gamma_y$ by the $\xi$-dependent
plateau values $\gamma_p$, determined by fitting $\gamma_y(h)$ to an
exponential decay with an offset. The data from the three
concentrations collapse onto a single curve (Fig.~\ref{rescale}B). 

We conclude that the nonlinear rheology of collagen is determined by
the interplay between the macroscopic material thickness and the mesh
size. We do not claim that the mesh size determines the network
rheology; this is most likely determined by some complex interplay of
the network structure and topology. In addition, we note that the range
of gaps investigated directly corresponds to the radial gap separation
in many cone-plate tool geometries, and is equivalent to length scales
of many important biological processes.

The observed size dependence of the nonlinear rheology indicates that
the material response is non-uniform on scales significantly larger
than the mesh size, and is therefore inconsistent with models that
attribute strain stiffening to the affine deformation of networks of
semi-flexible filaments.  Collagen fibers are relatively stiff, and
therefore more likely described by athermal models that attribute
strain stiffening to non-affine filament stretching
\cite{heuss_frey_07}.  The non-affine length scale can be
significantly larger than the mesh size \cite{lielig_prl_07}, but to
our knowledge the implications for system size dependence has not been
investigated.  Alternatively, heterogeneous localization of strain
  within the network that is either uniformly distributed through the
  gel or localized near the boundaries could produce size dependent
  effects.  Reflectance images do not show any measurable variation in
  collagen structure near the boundary, but more subtle boundary
  effects may also play a role.  Finally, the long persistence length
  of the collagen fibers may directly contribute to a finite size
  effect. Whatever the physical mechanism, the size-dependent
rheology extends to important length scales for a range of fundamental
studies and applications.
\vspace*{-17pt}
\noindent \section*{Acknowledgements}
\noindent 
We thank P.~Janmey, F.~MacKintosh and Andreas Bausch for helpful
discussions and W.~Rosoff and D.~Koch for the collagen
preparation. This was funded by the NSF through grant DMR-0804782 and
from the AFOSR through grant FA9550-07-1-0130.

%\bibliographystyle{nature} 
%\bibliography{grants}

\vspace*{-12pt}

\begin{thebibliography}{10}

\bibitem{gaill_biorheo_93}
Djabourov, M., Lechaire, J.-P., and Gaill, F. 1993.
\newblock Structure and rheology of gelatin and collagen gels.
\newblock {\em Biorheology}. 30:191--205.

\bibitem{weitz_04}
Gardel, M., Shin, J., MacKintosh, F., Mahadevan, L., Matsudaira, P., and Weitz,
  D. 2004.
\newblock Elastic behavior of cross-linked and bundled actin networks.
\newblock {\em Science}. 304:1301--1305.

\bibitem{janmey_nature_05}
Storm, C., Pastore, J., MacKintosh, F., Lubensky, T., and Janmey, P. 2005.
\newblock Nonlinear elasticity in biological gels.
\newblock {\em Nature}. 435:191--194.

\bibitem{liu_prl_07}
Liu, J., Koenderink, G.~H., Kasza, K.~E., MacKintosh, F.~C., and
Weitz, D.~A. 2007.
\newblock Visualizing the strain field in semiflexible polymer networks: Strain
  fluctuations and nonlinear rheology of $f$-actin gels.
\newblock {\em Phys. Rev. Lett.} 98:198304.

\bibitem{fletcher_nat_07}
Chaudhuri, O., Parekh, S.~H., and Fletcher, D.~A. 2007.
\newblock Reversible stress softening of actin networks.
\newblock {\em Nature}. 445:295--298.

\bibitem{janmey_prl_95}
MacKintosh, F.~C., K\"as, J., and Janmey, P.~A. 1995.
\newblock Elasticity of semiflexible biopolymer networks.
\newblock {\em Phys. Rev. Lett.} 75:4425--4428.

\bibitem{frey_prl_03}
Wilhelm, J. and Frey, E. 2003.
\newblock Elasticity of stiff polymer networks.
\newblock {\em Phys. Rev. Lett.} 91:108103.

\bibitem{head_pre_03}
Head, D.~A., Levine, A.~J., and MacKintosh, F.~C. 2003.
\newblock Distinct regimes of elastic response and deformation modes of
  cross-linked cytoskeletal and semiflexible polymer networks.
\newblock {\em Phys. Rev. E}. 68:061907.

\bibitem{frey_pre_07}
Heussinger, C., Schaefer, B., and Frey, E. 2007.
\newblock Nonaffine rubber elasticity for stiff polymer networks.
\newblock {\em Phys. Rev. E}. 76:031906.

\bibitem{onck_prl_05}
Onck, P.~R., Koeman, T., van Dillen, T., and van~der Giessen, E. 2005.
\newblock Alternative explanation of stiffening in cross-linked semiflexible
  networks.
\newblock {\em Phys. Rev. Lett.}  95:178102.

\bibitem{stein_micro_08}
Stein, A.~M., Vader, D.~A., Jawerth, L.~M., Weitz, D.~A., and Sander,
L.~M. 2008.
\newblock An algorithm for extracting the network geometry of three-dimensional
  collagen gels.
\newblock {\em Journal of Microscopy}. 232:463 -- 475.

\bibitem{kaufman_biophys_09}
Yang, Y.-L., Leone, L., and Kaufman, L. 2009.
\newblock Elastic moduli of collagen gels can be predicted from two-dimensional
  confocal microscopy.
\newblock {\em Biophysical Journal}. 97:2051--2060.

\bibitem{sackmann_biophys_03}
Forgacs, G., Newman, S., Hinner, B., Maier, C., and Sackmann, E. 2003.
\newblock Assembly of collagen matrices as a phase transition revealed by
  structural and rheologic studies.
\newblock {\em Biophysical Journal}. 84:1272--1280.

\bibitem{kaufman_05}
Kaufman, L., Brangwynne, C., Kasza, K., Filippidi, E., Gordon, V., Deisboeck,
  T., and Weitz, D. 2005.
\newblock Glioma expansion in collagen i matrices: Analyzing collagen
  concentration-dependent growth and motility patterns.
\newblock {\em Biophysical Journal}. 89:635--650.

\bibitem{janmey_natmat07}
Janmey, P.~A., Mccormick, M.~E., Rammensee, S., Leight, J.~L., Georges, P.~C.,
  and Mackintosh, F.~C. 2007.
\newblock Negative normal stress in semiflexible biopolymer gels.
\newblock {\em Nature Materials}. 6:48--51.

\bibitem{mackintosh_09}
Kang, H., Wen, Q., Janmey, P., Tang, J., Conti, E., and MacKintosh,
F. 2009.
\newblock Nonlinear elasticity of stiff filament networks: Strain stiffening,
  negative normal stress, and filament alignment in fibrin gels.
\newblock {\em Journal of Physical Chemistry B}. 113:3799--3805.

\bibitem{vader_plosone_09}
Vader, D., Kabla, A., Weitz, D., and Mahadevan, L. 2009.
\newblock Strain-induced alignment in collagen gels.
\newblock {\em PLoS ONE}. 4:e5902.

\bibitem{fabry_08}
Mierke, C.~T., R{\"o}sel, D., Fabry, B., and Br{\'a}bek, J. 2008.
\newblock Contractile forces in tumor cell migration.
\newblock {\em European Journal of Cell Biology}. 87:669--676.

\bibitem{taylor_08}
Ott, H.~C., Matthiesen, T.~S., Goh, S.-K., Black, L.~D., Kren, S.~M., Netof,
  T.~I., and Taylor, D.~A. 2008.
\newblock Perfusion-decellularized matrix: using nature's platform to engineer
  a bioartificial heart.
\newblock {\em Nature Medicine}. 14:213--221.

\bibitem{margaret_08}
Sabass, B., Gardel, M.~L., Waterman, C.~M., and Schwarz, U.~S. 2008.
\newblock High resolution traction force microscopy based on experimental and
  computational advances.
\newblock {\em Biophysical Journal}. 94:207 -- 220.

\bibitem{wang_02}
Beningo, K., Lo, C.-M., and Wang, Y.-L. 2002.
\newblock Flexible polyacrylamide substrata for the analysis of mechanical
  interactions at cell-substratum adhesions.
\newblock {\em Methods in Cell Biology}. 69:325--339.

\bibitem{wong_03}
Gaudet, C., Marganski, W., Kim, S., Brown, C., Gunderia, V., Dembo, M., and
  Wong, J. 2003.
\newblock Influence of type i collagen surface density on fibroblast spreading,
  motility, and contractility.
\newblock {\em Biophysical Journal}. 85:3329--3335.

\bibitem{heuss_frey_07}
Heussinger, C. and Frey, E. 2007.
\newblock Role of architecture in the elastic response of semiflexible polymer
  and fiber networks.
\newblock {\em Phys. Rev. E}. 75:011917.

\bibitem{lielig_prl_07}
Lieleg, O., Claessens, M. M. A.~E., Heussinger, C., Frey, E., and
Bausch, A.~R. 2007.
\newblock Mechanics of bundled semiflexible polymer networks.
\newblock {\em Phys. Rev. Lett.} 99:088102.

\end{thebibliography}

%\end{multicols}
\end{document}